\documentclass[aps,superscriptaddress,showpacs,nofootinbib,onecolumn]{revtex4}
\usepackage{bm}
\usepackage{xcolor}
\usepackage{graphicx,color}
\usepackage{amsmath}
\usepackage{amssymb}

\newcommand{\be}{\begin{equation}}
\newcommand{\ee}{\end{equation}}
\newcommand{\bea}{\begin{eqnarray}}
\newcommand{\eea}{\end{eqnarray}}

\begin{document}

\title{Chaotic measure of the transition between two- and three-dimensional turbulence}
 
\author{Daniel Clark}
\author{Andres Armua}%
\author{Calum Freeman}
\author{Daniel J. Brener}
 \author{Arjun Berera}

 \affiliation{%
 School of Physics and Astronomy, University of Edinburgh, JCMB,
\\King’s Buildings, Peter Guthrie Tait Road EH9 3FD, Edinburgh, United Kingdom.
}%

\date{\today}

\begin{abstract}

Using direct numerical simulation, we study the behavior of the maximal Lyapunov exponent in thin-layer turbulence, where one dimension of the system is constrained geometrically. Such systems are known to exhibit transitions from fully three dimensional turbulence through a mixed two- and three-dimensional phenomenology state and then onto fully two-dimensional dynamics. We find a discontinuous jump in the Lyapunov exponent at this second transition, implying the predictability of such systems can change dramatically. Such transitions are seen in a number of different turbulent systems, for example those undergoing strong rotation, hence these results may be relevant for the predictability of complicated real world flows. The Lyapunov exponent is found to provide a particularly clear measure of the transition to two-dimensional dynamics. Finally, the application of these results to atmospheric predictability with regards to high-resolution modeling is examined.

\end{abstract}

\maketitle

\section{Introduction}
The dynamical behavior of turbulent fluid flows is known to be vastly different in two and three dimensions. The three dimensional case is characterized by a forward cascade of energy from large to small scales separated by an inertial range where the energy spectrum takes the form $E(k) \sim k^{-5/3}$. In the two-dimensional case, the existence of a second quadratic invariant, the enstrophy, leads to a dual cascade scenario: an inverse cascade of energy from small to large scales and a direct enstrophy cascade from large to small scales. These cascades exhibit scaling regions of $E(k) \sim k^{-5/3}$ and $E(k) \sim k^{-3}$, respectively. Much of our understanding of turbulence in three dimensions can be attributed to Kolmogorov \cite{kolmogorov1941local, kolmogorov1962refinement}, whilst in two dimensions the groundwork was laid by Kraichnan \cite{kraich67}.

Despite these differences, there is a growing body of evidence that two- and three-dimensional turbulent dynamics can co-exist under certain circumstances. Perhaps the first demonstration of this was in measurements of the energy spectrum in the Earth's atmosphere \cite{nastrom}, in which the data was interpreted as showing both forward enstrophy and energy cascades. One possible explanation is that the geometry of the atmosphere is such that the vertical direction is constrained compared to the other two, with the result that above a certain length-scale the system is effectively two-dimensional. This situation is often referred to as thin-layer turbulence. The co-existence of two- and three-dimensional phenomenology, that is both forward and inverse energy cascades, has been observed in both experimental and numerical studies of thin-layers \cite{xia, byrne, izakov, young, celan, sesh, benav, biferale1, split, van2019}. In the numerical investigations, it was found that by reducing the thickness of the fluid layer the system transitions from fully three dimensional behavior to mixed two- and three-dimensional dynamics and then onto purely two-dimensional. Such transitions are not restricted to thin-layer turbulent systems; they have also been seen in turbulence undergoing rotation, exhibiting stratification, those under the influence of strong magnetic fields and in axis-symmetric flows \cite{smith, smith1, sozza, alexak, buzz, pest, bos}. For a more comprehensive review of such systems and cascade behavior see \cite{alexrev}.

These prior studies into this transition all employed the standard statistical approach to turbulence \cite{batchelor1953theory}, in which the properties of the flow under a suitable averaging procedure are studied. However, it is also possible to exploit the deterministic chaos exhibited by turbulent flows \cite{lorenz1963deterministic, ruelle1971nature, ott2002chaos, bohr2005dynamical} to investigate their behavior. As could be predicted from the differences in dynamical behavior across dimensions, the chaotic properties are also vastly different when comparing two and three dimensions. In particular, the scaling behavior of the maximal Lyapunov exponent, Kolmogorov-Sinai entropy and attractor dimension (see \cite{ott2002chaos} for definitions) in three dimensions was determined entirely by the Reynolds number of the flow \cite{berera2018chaotic, boffetta2017chaos, mohan,  berera2019info}. In two dimensions, for the case of the entropy and attractor dimension, this scaling was found to be more complicated and non-universal, being influenced by the system size and forcing length scale \cite{clark2020chaos}. The strong contrast between these two cases then suggests that these chaotic properties may be utilized in the study of this transitional behavior.

The use of the chaotic properties of a system in the study of phase transitions has seen a small amount of attention in the critical phenomena literature \cite{butera87, cainai97, barre01}. In such studies, it was found that the maximal Lyapunov exponent could be used as an indicator of a phase transition, showing differing behavior either side of a critical point. This, combined with the aforementioned drastic differences in the scaling behavior of chaotic properties of turbulent flows in two versus three dimensions, suggests the maximal exponent might provide a useful alternative viewpoint in the study of this transition in thin-layer flows. Furthermore, the Lyapunov exponent measured in numerical simulations of turbulent fluid flow is found to be a remarkably stable quantity, particularly against the effects of numerical resolution \cite{ho}. As such, it may be expected to be a robust measure of the transitional behavior seen in thin-layer turbulence. Finally, the Lyapunov exponent gives a measure of the predictability of a system. Given the observation of transitional behaviour seen in the Earth's atmosphere understanding how the predictability of thin-layer turbulence changes across such transitions may then provide important information for the wider study of atmospheric predictability.

\section{Problem set-up}
In this work, we study the transition between two- and three-dimensional phenomenology in thin-layer turbulence via measurement of the maximal Lyapunov exponent in direct numerical simulations (DNS) of the incompressible Navier-Stokes equations \begin{equation}
\begin{split}
\partial_t \bm{u} + \bm{u}\cdot \bm{\nabla}\bm{u} = -\bm{\nabla}P &+ \nu \nabla^2 \bm{u} +\mu \nabla^{-2} \bm{u} + \bm{f}, \\ \bm{\nabla} \cdot \bm{u} &= 0.
\end{split}
\end{equation} In the above: $\bm{u}(\bm{x},t)$ is the velocity field, $P(\bm{x},t)$ the pressure field, $\bm{f}(\bm{x},t)$ is an external force used to sustain the flow and $\nu$ is the kinematic viscosity. To avoid the formation of a large scale condensate as a result of the inverse energy cascade, we also include a hypo-viscous term with hypo-viscosity, $\mu$, which removes energy at the large scales. This is an important addition, as such a condensate is a form of self-organization, which can cause a reduction of chaos in the flow. In all cases $\mu$ is set such that a condensate is unable to form. We employ the standard pseudo-spectral method with full de-aliasing using the two-thirds rule. Our simulations are performed in a fully periodic box with side lengths $L\times L \times H$, with $L=2\pi$ and $H < L$. Throughout we will consider the side of length $H$ to be in the $z$ direction. To facilitate comparison with previous studies,  our external forcing function acts only on the $x$ and $y$ components of the velocity field and has the form $\bm{f} = (-\partial_y \phi, \partial_x \phi, 0)$ such that it is solenoidal. The scalar field $\phi(\bm{x},t)$ is stochastic and delta correlated in time, which ensures that on average the energy injection rate is $\varepsilon$, which can be set by the amplitude of the forcing. Additionally, the forcing is concentrated in Fourier space on modes with magnitude $k_f \simeq 2\pi/l_f$. The initial conditions of the flow are such that the field is near zero, with the small amount of energy spread across a wide range of length scales. As such, the flow is essentially generated by the stochastic forcing. We maintain an even grid spacing in physical space, therefore, upon reducing $H$ we also reduce the total number of grid points needed in the vertical direction. In Fourier space this leads to a larger spacing between modes in the vertical direction than in the horizontal directions.

As in \cite{van2019}, we find the system is described  by a number of non-dimensional parameters. The first is the Reynolds number defined at the energy injection scale as \begin{equation}\label{re}
\mathrm{Re} = \frac{\varepsilon^{\frac{1}{3}} l_{f}^{\frac{4}{3}}}{\nu}.
\end{equation} We also have the ratio of the forcing length scale and the side of length $H$ defined as $Q = l_f / H$ and the aspect ratio of the system given by $A = H/L$. Of these two quantities, It has been observed that $Q$ is the more important parameter for determining the transition points of the system \cite{benav,van2019}, thus, we formulate our results in terms of $Q$. For $Q$ much less than 1, at the length scale where energy is injected the system is fully three dimensional, and as such we expect three dimensional phenomenology to dominate. For $Q$ much greater than 1 the system is expected to appear two-dimensional. In between these two extremes it is observed that both two- and three-dimensional behavior coexist in the flow. Indeed, in \cite{benav} using a severe Galerkin truncation in the vertical direction, an Re independent critical value of $Q$ was found, above which the flow transitions from three dimensional behavior to mixed phenomenology. Additionally, a second critical value at which point the system moves to two-dimensional behavior was found, although this point was found to have an Re dependence. Our simulations span the ranges Re $\approx 90 - 1200$ and $Q \approx 0.1 - 16$ with the forcing length scale either $k_f = 4, 8$. Finally, the number of grid-points used in the horizontal directions varied from $128-1024$, such that the simulations remained well resolved.  

\begin{figure}
    \includegraphics[width=\columnwidth]{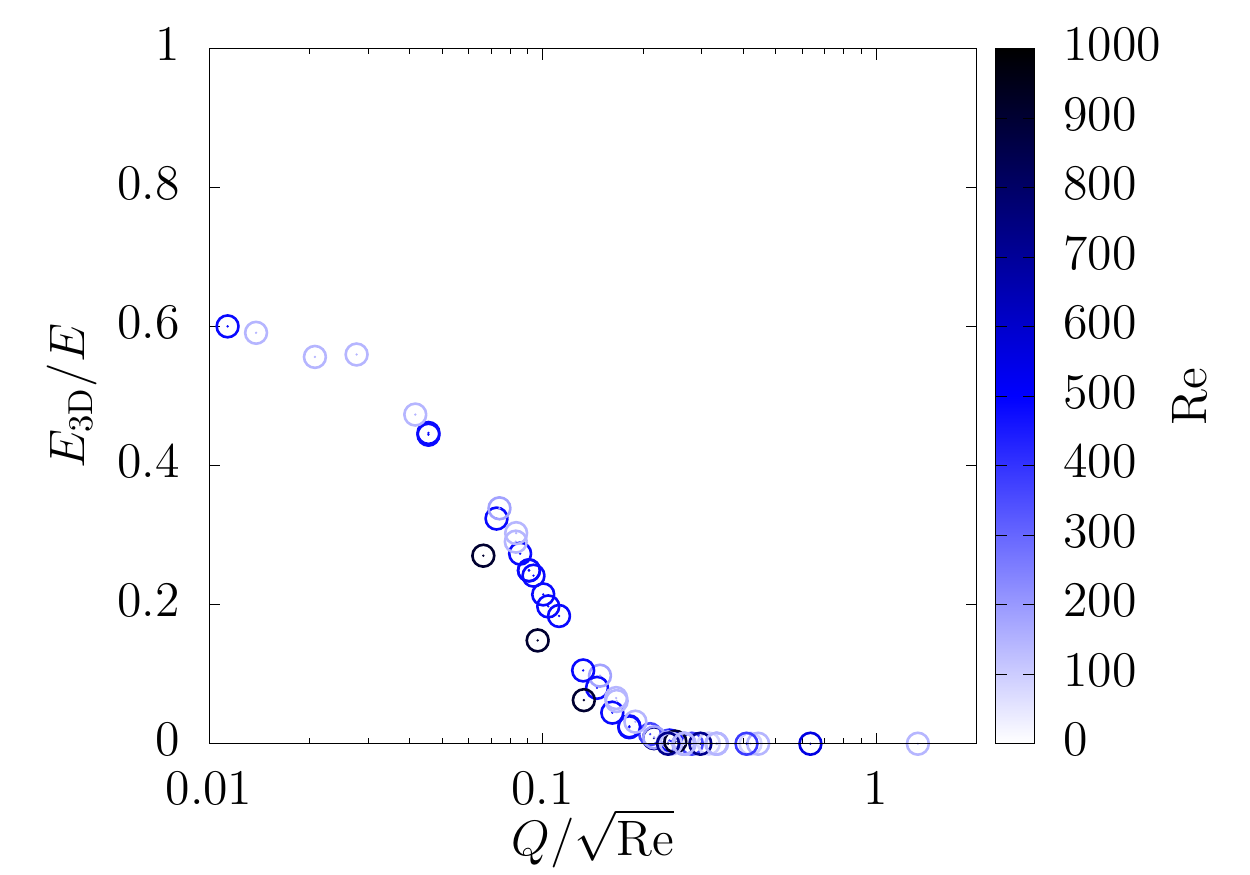}
    \caption{\label{e3de}Ratio of average energy in the three dimensional modes to the average total energy across a range of $Q$ and Re values. Color gradient used to indicate Re value, becoming darker as Re increases.}
\end{figure} 

We will focus here on this second critical value, denoted as $Q_{\mathrm{2D}}(\mathrm{Re})$. In \cite{benav} it was shown that by considering the interplay between the layer thickness $H$ and the shearing force driving 3D instabilities in the flow, this critical thickness should behave as \begin{equation}
Q_{\mathrm{2D}}(\mathrm{Re}) \propto \sqrt{\mathrm{Re}},
\end{equation} see also \cite{gallet} for further information. Before presenting results for the maximal Lyapunov exponent in thin-layer turbulence, we will first establish an approximate value in terms of $Q/\sqrt{\mathrm{Re}}$ at which the transition to two-dimensional dynamics occurs using a standard indicator. We will consider the velocity field to be decomposed into two- and three-dimensional parts \textit{i.e.} \begin{equation}
\begin{split}
\bm{u}(\bm{k},t) &= \bm{u}_{\mathrm{2D}}(\bm{k},t) + \bm{u}_{\mathrm{3D}}(\bm{k},t) \\ &= \bm{u}(\bm{k}:k_z = 0,t) + \bm{u}(\bm{k}:k_z\neq0,t),
\end{split}
\end{equation} such that the two-dimensional part is composed of all modes with vertical wavenumber, $k_z = 0$. Using this decomposition, the total energy of the flow also becomes split into two- and three-dimensional parts \begin{equation}
E(t) = E_{\mathrm{2D}}(t) + E_{\mathrm{3D}}(t).
\end{equation} At the point $Q_{\mathrm{2D}}(\mathrm{Re})$, we expect the three dimensional energy to vanish. As such, we consider the ratio of the averaged three dimensional energy to the averaged total energy. In Fig. \ref{e3de} we plot this ratio for a range of $Q$ and Re values. Here, we see a common curve across all Re values, with the possible exception of only the highest Re values. This is likely explained by the transition from three dimensional behavior to mixed dynamics becoming Re independent at high enough Re, as found in \cite{van2019}. For all cases we find the transition point $Q_{\mathrm{2D}}(\mathrm{Re})$ to occur at $Q/\sqrt{\mathrm{Re}} \approx 0.25$. This is consistent with what was found in the pre-condensate phase of the simulations in \cite{van2019}. It should be noted that the value of $Q_{\mathrm{2D}}(\mathrm{Re})$ may be influenced by the form of forcing employed. In our case, the forcing is fully two-dimensional, however, it has been found that when using a three dimensional force the transition point is altered \cite{alex2}.

\section{Lyapunov exponents in fluid turbulence}

Having established a value for $Q_{\mathrm{2D}}(\mathrm{Re})$, we now consider the maximal Lyapunov exponent. This exponent gives, to leading order, the rate of divergence of trajectories in the state space of the system. This state space in our simulations is of very high dimension, equal to the number of Fourier modes retained. Systems with a positive maximal Lyapunov exponent are said to exhibit deterministic chaos, a state characterized by a extreme sensitivity to initial conditions. Turbulent fluid flow is known to be deterministically chaotic and, given the ubiquity of turbulence in nature, this has implications for the predictability of real world phenomena. This work is motivated in part by atmospheric predictability, where the chaotic nature of turbulence, coupled with finite experimental resolution, leads to small measurement errors growing exponentially in time. As a result, if weather forecasts are to be accurate within a given error tolerance there is a finite predictability time before the error will grow to exceed any tolerance. This predictability time is determined by the maximal Lyapunov exponent, at least for small scale weather phenomena. For larger scale climate forecasting it is possible for the predictability time to exceed that given by the Lyapunov exponent. We return to this point in Sec. \ref{conc}.

Numerically, the maximal Lyapunov exponent is calculated by considering two distinct velocity fields, one perturbed slightly from the other once a steady state has been reached. More explicitly, we consider a reference field $\bm{u}_1$ and a perturbed field, $\bm{u}_2$, which at the perturbation time $t_0$ is defined as \begin{equation}
\bm{u}_2(t_0) = \bm{u}_1(t_0) + \bm{\delta_0},
\end{equation} in which $\bm{\delta_0}$ is a Gaussian random velocity field with zero mean. The variance of the perturbation field is chosen such that the intitial seperation, $\Delta$, between the two velocity fields can be considered infinitesimal, i.e. $|\bm{\delta_0}| \equiv \Delta \ll U $, where $U$ is the RMS velocity. The use of an infinitesimal perturbation can be considered to model the effect of finite experimental measurement resolution, with the reference field representing the true system, and the perturbed field representing the what is measured experimentally. Both fields are then evolved concurrently according to the Navier-Stokes equations. Whilst the difference between the fields remains small, it is found to grow exponentially in time with a rate given by the maximal Lyapunov exponent. To obtain more detailed statistical information for this exponent, the difference between the two fields is re-scaled to the initial value $\Delta$ at periodic intervals of time of length $\Delta t$ \begin{equation}
\bm{u}_2(t_0 + \Delta t) = \bm{u}_1(t_0 + \Delta t) + \Delta \frac{\bm{\delta}(t_0 + \Delta t)}{|\bm{\delta}(t_0 + \Delta t)|},
\end{equation} in which $\bm{\delta}(t) = \bm{u}_2(t) - \bm{u}_1(t)$. The finite time Lyapunov exponent (FTLE) is then defined as \begin{equation}
\gamma(\Delta t) = \frac{1}{\Delta t}\ln \left(\frac{|\boldsymbol{\delta}(\Delta t)|}{\Delta} \right),
\end{equation} which, when averaged over many iterations, gives the maximal Lyapunov exponent $\lambda = \langle \gamma(\Delta t) \rangle$. For more detailed descriptions of numerically computing Lyapunov exponents see \cite{benettin1980lyapunov}. Importantly, if, as in this work, a stochastic forcing is used, the force should only be randomly generated once per iteration and the same force applied to both the reference and perturbed fields. If a new random force is generated for each field this acts like a new perturbation at each iteration and destroys the exponential growth of separation between the fields.

\begin{figure}
    \includegraphics[width=\columnwidth]{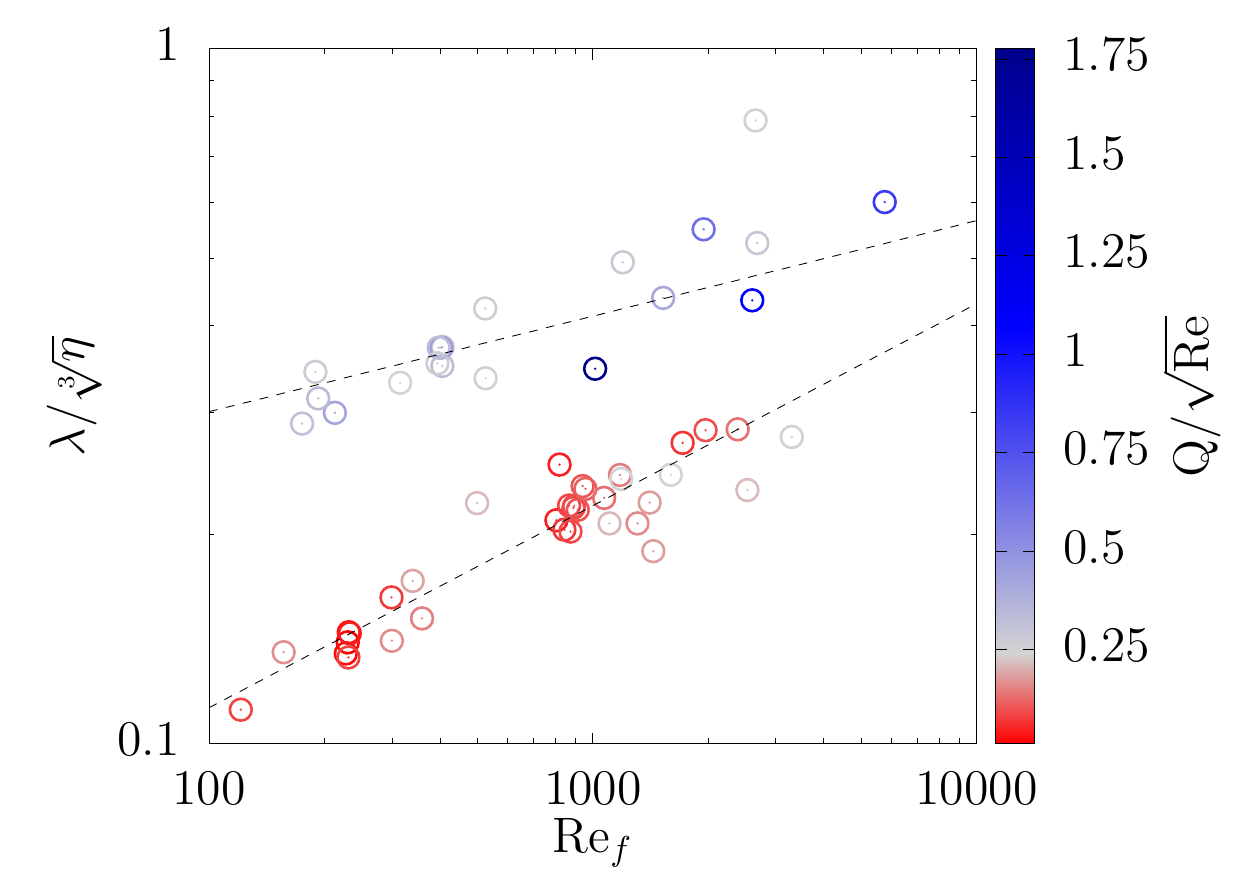}
    \caption{\label{lyapRe} Maximal Lyapunov exponent, $\lambda$, scaled by the enstrophy dissipation rate timescale against Re. Color gradient is set such that approaching the transition at $Q/\sqrt{\mathrm{Re}}$ from either side results in a lighter color. Additionally points below the transition become more red whilst above they become more blue. Lower dashed line has a gradient of 0.29 whilst higher dashed line has a gradient of 0.14, and these give the corresponding Re$_E$ scaling exponents.}
\end{figure} 

\begin{figure}
    \includegraphics[width=\columnwidth]{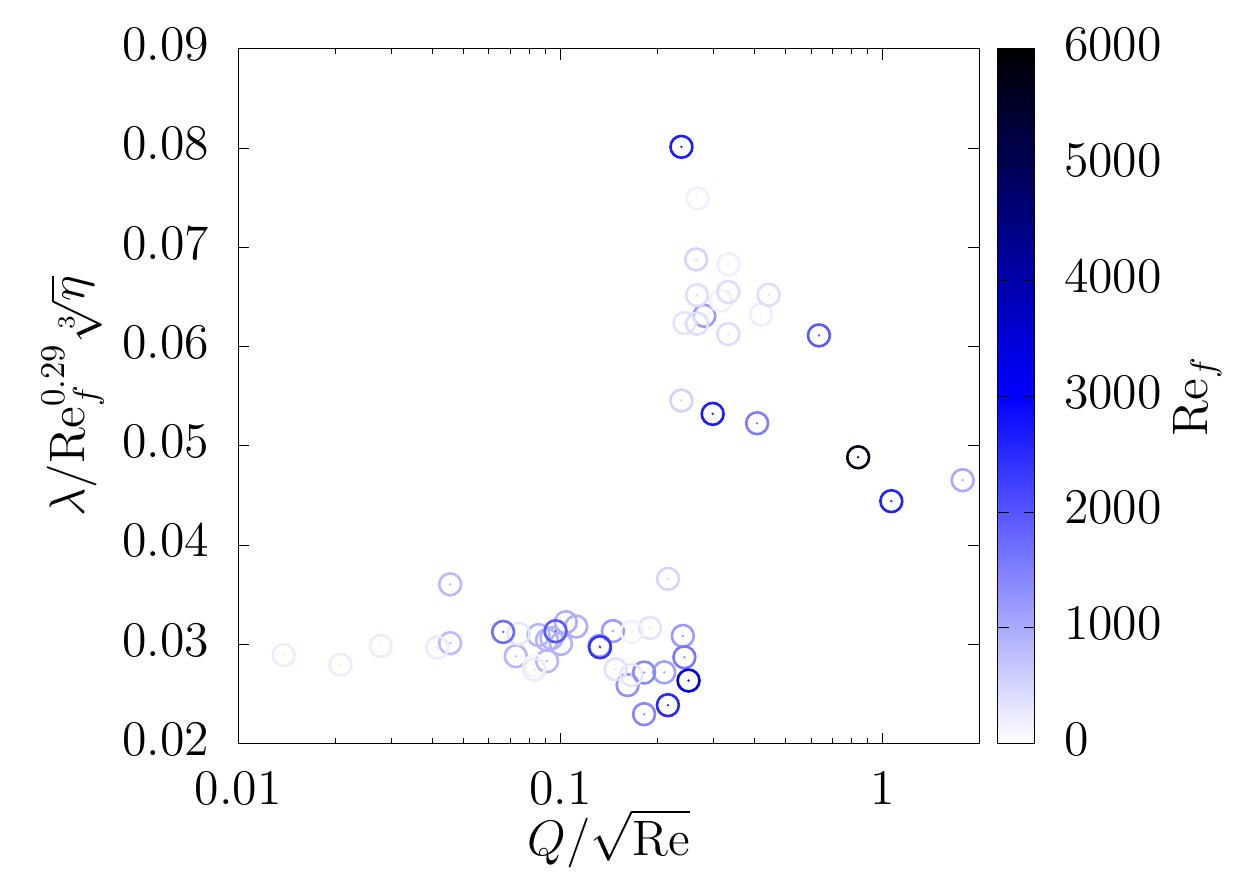}
    \caption{\label{lyapQ} Maximal Lyapunov exponent, $\lambda$, scaled by the Reynolds number and enstrophy dissipation rate against $Q/\sqrt{\mathrm{Re}}$. Color gradient used to indicate Re$_f$ value, becoming darker as Re$_f$ increases. An empirically determined exponent is used in scaling the maximal exponent. This is to highlight the discontinuous transition.}
\end{figure}

The expected scaling behavior of $\lambda$ in both two- and three-dimensional turbulence can be estimated on dimensional grounds by assuming it will be determined by the inverse of the smallest timescale of the flow. In homogeneous and isotropic three dimensional turbulence this is given by the Kolmogorov time scale, which then implies \cite{ruelle1979microscopic, crisanti1993predictability} \begin{equation}
\lambda_{\mathrm{3D}} \sim \tau^{-1} \sim \frac{1}{T}\sqrt{\mathrm{Re}},
\end{equation} where $\tau$ is the Kolmogorov time and $T$ is the large eddy turnover time. Following similar arguments in two dimensions results in \cite{ohkitani} \begin{equation}
\lambda_{\mathrm{2D}} \sim \sqrt[3]{\eta},
\end{equation} in which $\eta$ is the enstrophy dissipation rate. In the two-dimensional case there is also the possibility of logarithmic dependence on Re due to logarithmic corrections to the energy spectrum in the enstrophy scaling range \cite{kraich71}. It is not clear which of these dimensional estimates should be used in thin-layer turbulence . Furthermore, there is no theory in the literature to guide this choice. 

\section{Results}

To investigate the transition from the viewpoint of predictability, we consider the Re dependence of $\lambda$. Note that here, similarly to in \cite{benav}, we use a second forcing scale Reynolds number defined  as \begin{equation}
\mathrm{Re}_f = \frac{l_f \sqrt{E}}{\nu}.
\end{equation} The reason for second definition is that the Re defined in Eq. \ref{re} does not contain any information about the dynamical properties of the underlying flow, making determination of scaling exponents difficult. Note, that we continue to use Re as defined in Eq. \ref{re} when discussing the location of $Q_{\mathrm{2D}}(\mathrm{Re})$ to facilitate comparison with the literature. In Fig. \ref{lyapRe} we clearly observe two distinct scaling laws, one for points below $Q_{\mathrm{2D}}(\mathrm{Re})$ and another for those above. It should be noted that, in contrast with these previous studies in three dimensions, we non-dimensionalize $\lambda$ using the enstrophy dissipation rate timescale. This is justified by the relationship between enstrophy production and velocity derivative skewness which, if we assume K41 holds, gives this timescale the same Re scaling as $\tau$. The scaling exponent below the critical point is in good agreement with the $\lambda \tau$ scaling shown in \cite{berera2018chaotic} for purely three dimensional turbulence. In a study of two-dimensional turbulence \cite{clark2020chaos} the Re dependence of $\lambda$ was found to be $\lambda\sim \mathrm{Re}^{0.16}$ which is in line with what we find for beyond $Q_{\mathrm{2D}}(\mathrm{Re})$. Fig. \ref{lyapRe} highlights the possibility of an increase in Re, seemingly paradoxically, causing an increase in predictability as the system moves from the inverse cascade branch to the bidirectional cascade branch. Given that the value of $Q/\sqrt{\mathrm{Re}}$ at which this change in scaling occurs is the same as for the energy indicator and that seen in the literature, it is clear that the Lyapunov exponent provides a robust measure of the transition.

In Fig. \ref{lyapQ} we show $\lambda$ re-scaled by both Re and $\eta$ for a range of $Q$ and Re values. The power of Re$_f$ chosen corresponds to the scaling exponent for points below the transition point in Fig. \ref{lyapRe}. This scaling gives an approximately constant value for points below $Q/\sqrt{\mathrm{Re}} \approx 0.25$. At and above this point, we observe what appears to be a discontinuous jump as the flow becomes two-dimensional and the scaling behaviour is changed. 

Notably, in both Fig. \ref{lyapRe} and \ref{lyapQ} there is no indication of a first transition from three dimensional to mixed dynamics. This suggests the leading chaotic properties of the flow remain effectively fully three dimensional until the point $Q_{\mathrm{2D}}(\mathrm{Re})$. This is in agreement with the idea that the maximal Lyapunov exponent should be related to the shortest timescale of the flow. In both the three dimensional and mixed states, a forward cascade of energy to the smallest scales is seen, only vanishing when we pass $Q_{\mathrm{2D}}(\mathrm{Re})$.  Physically, we can understand this behavior by considering the cascades and triadic interactions involved at each stage. As we increase $Q$, the triads corresponding to three dimensional dynamics are progressively removed. Upon reaching the transition point, a critical proportion of these triads have been lost, ending the forward energy cascade and rendering the flow two-dimensional. Unlike for $E_{\mathrm{3D}}/E$, which decreases continuously as the forward cascade region is reduced, if $\varepsilon$ and $\nu$ are fixed, then as $Q$ is increased, and as long as a forward cascade exists, $\lambda$ will have the same value before changing discontinuously when no forward cascade remains. A similar discontinuous transition was observed in \cite{sahoo}, where the energy cascade was reversed by altering the weighting of certain triadic interactions between helical modes. See also the appendix of \cite{biferale1} for further discussion of this triadic interpretation.

It is also possible to view the transition through a less abstract physical interpretation, albeit one that is intimately connected with the triadic interpretation. It is well known that in two dimensions the vorticity equation has no vortex stretching term \cite{DavidsonBook}. As a result, since we consider the steady state case, the enstrophy dissipation timescale in our simulations in the $Q/\sqrt{\mathrm{Re}} > 0.25$ regime is set entirely by the rate of enstrophy injection by the forcing. On the other side of the transition vortex stretching is possible, and thus additional enstrophy is produced by non-linear interactions. This then leads to a higher rate of enstrophy dissipation at steady state for an equal rate of enstrophy injection. Hence, in the $Q/\sqrt{\mathrm{Re}}< 0.25$ regime the maximal Lyapunov exponent is determined by the smallest scales of the flow. The transition can then be understood as a consequence of the effect of geometric confinement on vortex stretching. The predictability of flows on either side of the transition can then be vastly different: on the three-dimensional side it is governed by action at the smallest scales of the flow and by the macroscopic energy injection scale on the two-dimensional side. 

An important point to consider in the simulation results presented here is the influence of the form of geometric confinement used. It is clear that considering a periodic flow in a box with a varying height is a very artificial kind of confinement. In atmopsheric turbulence, flow confinement is typically seen as a result of stratification. As mentioned in the introduction, stratified flows are also known to exhibit transitions between two- and three-dimension turbulence, dependent on the degree of stratification. It would be interesting to study the predictability of the transition in these flows as a more representative approximation of atmospheric turbulence. It may be possible to make a connection between the Reynolds number dependent $Q$ criterion for the transition seen here and the Richardson stability criterion of stratified flows \cite{richardson}, particularly as both are related to vortex stretching.

An interesting area for investigation is then what happens to the behavior of the chaotic properties of the flow which depend on all the active degrees of freedom in the flow, for example the Kolmogorov-Sinai entropy and attractor dimension \cite{ott2002chaos}. These may show different transitional behavior and reveal further information about the properties of such transitions. However, their calculation is outwith the scope of this work.

Since $\lambda$ represents the exponential rate at which two initially close fields diverge from each other, it provides a measure of the predictability time of the flow. Fig. \ref{lyapQ} suggests that this predictability time will exhibit discontinuous jumps. This is of particular interest in real world thin-layer systems which are, in general, non-stationary. In such flows, as Re varies and the flow transitions from one set of dynamics to another, predictability may be drastically altered.

\begin{figure}
    \includegraphics[width=\columnwidth]{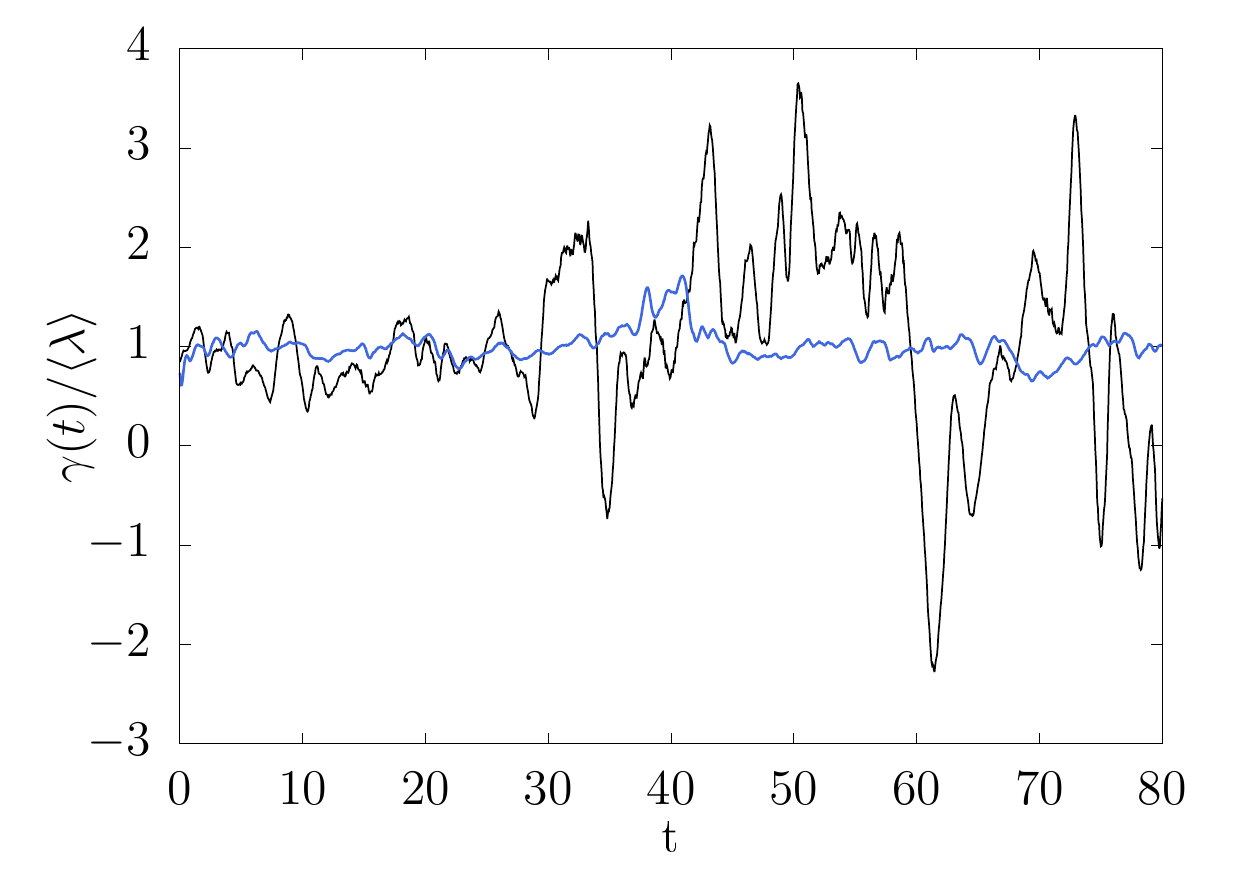}
    \caption{\label{lyapT} Time series for the finite time Lyapunov exponent, $\gamma(t)$ re-scaled by the mean value. Time is measured from the point the exponent has stabilized. We show a case far from $Q/\sqrt{\mathrm{Re}}\approx 0.25$ (blue) and one close to this point (black). These cases have similar Reynolds number values with Re $\approx 600$.} 
\end{figure}

Finally, we have also studied the temporal behavior of $\lambda$. In \cite{van2019}, as the point $Q_{\mathrm{2D}}(\mathrm{Re})$ was approached, intermittent bursts of three dimensional energy were observed and related to the idea of on-off intermittency in dynamical systems \cite{yama}. Such bursts should impact the behavior of the finite time Lyapunov exponents and would be expected to cause large fluctuations. Indeed, in Fig. \ref{lyapT} the case with $Q/\sqrt{\mathrm{Re}}\approx 0.25$ is seen to undergo large variations in time. As the cases shown are at comparable Re values then we can be relatively confident these fluctuations are caused by proximity to the transition. Although we only show two cases, this behavior is typical of points close to the transition point.

The appearance of large fluctuations as we approach the transition point is reminiscent of phase transitions in critical phenomena. It is then tempting to try to classify this transition from a bidirectional cascade to two-dimensional dynamics. Indeed, the abrupt change in behavior of the Lyapunov exponent at $Q_{\mathrm{2D}}(\mathrm{Re})$ suggests something similar to a first order phase transition may be occurring. Making a definitive statement on this issue will require further investigation and a wider range of the parameter space to be studied. 

\section{Concluding remarks}\label{conc}

Using Lyapunov exponents in systems of complexity to study phase transitions has received little attention in the literature. Therefore, their utilization in this work on fluid turbulence, and the clarity of the results achieved, suggests this method may be of particular use in extended non-equilibrium systems in general. A particular application of our results may well be found in the next generation of numerical weather prediction models. For systems with multiple timescales, the Lyapunov exponent is proportional to the smallest characteristic timescale, regardless of the size of the fluctuations in the different timescales. In the atmosphere, predictions can be made beyond the limit imposed by the Lyapunov timescale, which is associated with turbulence, as the predictability is imposed by the large scale dynamics \cite{lorenz1963deterministic,lorenz1969predictability}. However, within the last decade increases in computing power have allowed for large-eddy simulations to be nested within numerical weather prediction models, whereby three dimensional turbulence is resolved explicitly. These high-resolution simulations have important applications in many areas, such as particle transport dispersion modeling and wind turbine site profiling \cite{LES}. 

Furthermore, the advent of exascale computing will see operational global weather models run at far greater resolution ($<1$km), which will allow regional models to operate at scales where turbulent phenomena are explicitly resolved \cite{neumann_rs}. Therefore, our finding of a discontinuous change in predictability at $Q_{\mathrm{2D}}(\mathrm{Re})$ indicates that understanding the transition between two- and three-dimensional turbulent regimes in the atmosphere may be essential for determining predictability in different weather scenarios in these future high-resolution regional models. Forecast skill could be improved, particularly in severe convective thunderstorms, by more accurately resolving the atmosphere's transition from predominantly two to three-dimensional turbulent motion, which occurs in convection, as the error growth may change rapidly across this transition \cite{storm_example, resolving_stormy_clouds, resolving_stormy_clouds2, insights_judt}. Additionally, in-situ aircraft observations have shown that in the hurricane boundary layer, a height-dependent transition between two- and three-dimensional turbulence occurs, and that the large-scale hurricane vortex feeds directly from the small scales \cite{byrne}. Given this, our result that the transition from two- to three-dimensional dynamics is accompanied by a discontinuous change in the Lyapunov exponent means that correctly resolving this transition in hurricane models will be necessary for correctly predicting changes in intensity. 

Based on the results of this study into geometrically confined turbulence alone it is not yet possible to make any definitive claims regarding real-world atmospheric turbulence. However, the potential applications discussed in the preceding paragraphs suggest that further study of transitions in turbulence through the lens of predictability should be carried out. It is known that transitions between two- and three-dimensional turbulence occur in both rotating and stratified flows. Considering the transition in these cases from the point of view of triadic interactions presents a different picture than in the problem studied here. In our simulations, the flow is geometrically confined in an artificial way by varying the height of the domain, resulting in certain triads being removed entirely from the flow. This is not the case in rotating and stratified flows, where the effects of rotation and stratification will progressively damp certain triads. It is then possible that the predictability of these flows around the transition will not show the same behavior as seen in thin-layer turbulence. Given that these flows better approximate true atmospheric turbulence than the problem studied in this work, understanding their predictability will allow more concrete claims to be made regarding the potential applications discussed here.

A final caveat to consider when interpreting the findings presented here is the range of applicability of the Lyapunov exponent in numerical weather forecasting. In the atmosphere it is the large-scale dynamics that determine long-term predictability, hence, predictability is found beyond the timescale defined by the maximal exponent \cite{lorenz1969predictability}. However, for local, short-time scales, predictability is dominated by the non-linear dynamics of the system and thus the Lyapunov exponent is a useful measure. Hence, for the next generation of high resolution numerical weather prediction models, particularly those used in nowcasting, the transitions in predictability as measured via the maximal Lyapunov exponent may become important to accurately resolve. For a more in depth discussion of some of these points see \cite{weatherClimate} and \cite{timmermann_jin_2006}.

To summarize, we have studied the behavior of the maximal Lyapunov exponent in thin-layer turbulence through the use of direct numerical simulation. Using this exponent, we have measured the point at which the flow transitions from a bidirectional energy cascade to a purely inverse energy cascade. This point was found to occur at $Q/\sqrt{\mathrm{Re}} \approx 0.25$ when measured from Lyapunov exponent data, which is in agreement with the value obtained via more standard methods \cite{van2019}. The nature of the transition when viewed through the Lyapunov exponent is abrupt and discontinuous. As the maximal exponent is determined by the small scale features of the flow, it is not sensitive to the transition from a purely forward cascade to a bidirectional cascade. This suggests the short time predictability of such bidirectional cascade systems is as in three dimensions. However, near the transition to a purely inverse cascade the potentially discontinuous nature of the transition leaves the possibility for dramatic changes of the predictability time in this region. Our results demonstrate that, almost paradoxically, the predictability of a system can change discontinuously even when other quantities, such as the energy, vary smoothly. As such, studying this transition via the Lyapunov exponent provides a complementary approach, and highlights the importance of resolving these effects in future models of atmospheric predictability.

\begin{acknowledgments}
This work used the Cirrus UK National Tier-2 HPC Service at EPCC \cite{cirrus} funded by the University of Edinburgh and EPSRC (EP/P020267/1). This work has used resources from ARCHER \cite{archer} via the Director's Time budget. D.C. and A.A are supported by the University of Edinburgh. D.J.B is supported by the Carnegie trust. A.B. acknowledges partial funding from the U.K. Science and Technology Facilities Council.

\end{acknowledgments}

\end{document}